\documentclass{article}

\usepackage{graphicx}
\usepackage{amsmath}
\usepackage{authblk}
\usepackage{subfig}

\begin{document}

\title{A procedure for detecting hidden surface defects in a plate from real thermal data by means of active thermography}

\author[1]{Gabriele Inglese\thanks{gabriele@fi.iac.cnr.it}}

\author[2]{Roberto Olmi \thanks{r.olmi@ifac.cnr.it}}

\author[2]{Saverio Priori}

\affil[1]{IAC `M. Picone'' - CNR, Via Madonna del Piano 10, 50019 Sesto Fiorentino (Italy)}
\affil[2]{IFAC - CNR, Via Madonna del Piano 10, 50019 Sesto Fiorentino (Italy)}

\date{}

\maketitle

\noindent {\it Abstract} Let $\Omega_\epsilon$ be a metallic plate whose top inaccessible surface has been damaged by some chemical or mechanical agent. We heat the opposite side and collect a sequence of temperature maps $u^\epsilon$. Here, we construct a formal explicit approximation of the damage $\epsilon\theta$ by solving a nonlinear inverse problem for the heat equation in three steps: (i) smoothing of temperature maps, (ii) domain derivative of the temperature, (iii) thin plate approximation of the model and perturbation theory.

Our inversion formula is tested with realistic synthetic data and used in a real laboratory experiment.\\

\section{Introduction.}
Active thermography is a non-contact, non-destructive technique exploiting the information contained in the thermal contrast to gain knowledge about the integrity of a material structure. 

Since any material object is subject to aging, monitoring the deterioration is an essential task. In particular, corrosion of metallic structures poses a huge problem to the maintenance of industrial and public assets like steel bridges, chemical and nuclear power plants, pipelines and others. 

Nondestructive testing plays a crucial role in the effort of an early detection of the decay of metal structures. Applications of thermography have been reported, for example, in the diagnostics of aircraft structures \cite{SWCH93} \cite{CHS95}, in the integrity analysis of pipelines \cite{KYGWK14}, in the detection of cracks in steel bridges \cite{SIKMK14} and, in general, in the investigation of thick metallic structures \cite{GBM07} \cite{VC12}.\\

In this paper we process thermal data to evaluate inaccessible surface damages on a plate made of a heat-conducting material. In particular, we obtain an explicit approximation of the damage by means of perturbative methods.\\

For simplicity, we deal with a flat geometry. The undamaged plate is modeled by the parallelepiped $\Omega_0=\{ (x,y,z) \in (-\frac{L}{2},\frac{L}{2})\times (-\frac{L}{2},\frac{L}{2}) \times (0,a)\}$ ($0< a \ll L$).  Let $c\rho$ be the volumetric heat capacity ( $c$ is the specific heat capacity and $\rho$ is the density) and $\kappa$ be the thermal conductivity. The specimen is assumed homogeneous and isotropic so that $c$, $\rho$ and $\kappa$  are positive scalar constants. We refer to \cite{CJ59} for a complete background about the classical theory of heat.\\

Suppose that $\Omega_0$
divides an {\it outer} aggressive environment from our laboratory. Let  $S_{Top}=\{(x,y,a) \phantom{x} x,y \in [-\frac{L}{2},\frac{L}{2}] \}$ be the inaccessible face of $\Omega_0$ in contact with the environment while  $S_{Bot}=\{(x,y,0) \phantom{x} x,y \in [-\frac{L}{2},\frac{L}{2}] \}$ is the laboratory side.
We are able to heat the specimen $\Omega_0$ with a device
(a couple of $1000W$ spotlights in our case) that produces a heat flux $\phi(x,y)$ through $S_{Bot}$ for a time interval $(0,T_{max})$. In the meanwhile, the increase of surface temperature on $S_{Bot}$ is monitored by means of an infrared camera. A sequence of temperature maps is stored. 

Let $u^0(x,y,z,t)$ be the temperature at the point $(x,y,z) \in \Omega_0$ at time $t$ ({\it background temperature} at time $t$).\\

In this framework, we suppose that a damage, due to  chemical or mechanical aggression could appear on the top side. The damaged specimen is called $\Omega_{\epsilon}$. The temperature maps of the bottom side of $\Omega_{\epsilon}$ are  $u^\epsilon_k=u^\epsilon(x,y,0,t_k)$ for $0<t_1<t_2<...<t_N\le T_{max}$. Our goal is to identify the damage from the knowledge of the {\it thermal contrast} 
$$u^\epsilon(x,y,0,t_k)-u^0(x,y,0,t_k)$$
for $0<t_1<t_2<...<t_N\le T_{max}$. 

This procedure is called Stepped Heating Thermography (SHT: see for example \cite{Ma01} Section 9.3). A technical and  historical introduction to Thermal NonDestructive Testing is in the Preface of \cite{Ma01}. In  Table 1.2 of  \cite{Ma01} a detailed list of applications is reported. In particular, SHT is mainly used in the thermal testing at low conductivities or to check composite materials (carbon fiber-reinforced plastic, chips, adhesive). Here, we apply SHT to the NDT of plates with both high and low conductivity.\\ 

\noindent{\bf Nomenclature:}\\

\noindent {\it Parameters}\\

$\Omega_0$  undamaged plate

$\Omega_\epsilon$ damaged plate

$S_{Top}$  surface of $\Omega_0$ in contact with environment (inaccessible)

$S_{Bot}$  laboratory side of $\Omega_0$

$\epsilon$ depth of the damage

$\theta$   shape of the damage

$U^0$  initial and environmental temperature

$h$   heat transfer coefficient

$\phi$  heat flux density

$T_{max}$	time length of the experiment

$\kappa$   conductivity

$\rho$  density

$c$  specificific heat capacity

$\alpha=\frac{\kappa}{\rho c}$ diffusivity
\\

\noindent {\it Acronymos}\\

IBVP Initial Boundary Value Problem 

IP   Inverse Problem

TPA  Thin Plate Approximation

FEM  Finite Elements Method\\

\subsection{The Inverse Problem}\label{ssec:ip}

We have assumed that the only effect
of the external aggression to the specimen is the loss of an amount of matter so that $S_{Top}$ becomes an unflat surface (see for example \cite{BKW90}, \cite{BK05}, \cite{KSV94}). Furthermore, we suppose that a nonnegative function $\epsilon\theta$ describes the deviation of damaged $S_{Top}$ from  original plane. More precisely,
$$\Omega_{\epsilon}=\{(x,y,z): x,y \in (-\frac{L}{2},\frac{L}{2}) ; z<a-\epsilon\theta(x,y) \}.$$

\noindent We can formulate the inverse problem {\bf IP}:\\

\noindent {\bf IP} Detect and evaluate $\epsilon\theta$  from the knowledge of $\phi$ (controlled heat flux) and $u^\epsilon(x,y,0,t_k)-u^0(x,y,0,t_k) $ for $k=1, ... , N$ (maps of the thermal contrast).\\

We point out that the heat flux $\phi$ through $S_{Bot}$ is actually unknown because of factors like surface reflection of electromagnetic radiation and dispersion of heat in the air between lamps and plate. For this reason, in practice we must identify $\phi$ from the temperature maps taken with the infrared camera which, ultimately, are the only available data (see section \ref{ssec:phi}).\\

\noindent {\it Remarks about how we model damages}\\

\noindent  {\it 1. The defect is described by a function}. A surface defect is naturally modeled by means of a function
of two variables when it is originated by  mechanical collisions or some kind of corrosion (for example, uniform corrosion or erosion-corrosion). On the other hand, pitting corrosion or cracking  (see for example \cite{J96} (section 1.5)) must be modeled by means of a curve that is not the graph of a function (\cite{BVV10}).\\

\noindent  {\it 2. The defect is actually time-independent}. Although the formation of defects on $S_{Top}$ is a dynamical process, we model it by means of a time-independent function $\epsilon\theta$. It is correct as long as the time scale of damaging evolution is much larger than the observation time $T_{max}$.\\

\subsection{Details of the direct model}

The temperature in  $\Omega_{\epsilon}$ for  $t \in (0,T_{max}]$, fulfills the  heat equation
\begin{equation}\label{eq:01}
 u_t=\alpha \Delta u
\end{equation}
with  boundary conditions that account for energy exchanges between the specimen and the environment: 
\begin{equation}\label{eq:02}
\kappa u_n(\sigma,t)+h(u(\sigma,t)-U^0)-\phi(\sigma)=0
\end{equation}
where $\sigma \in \partial \Omega_\epsilon$. These are called Robin or third kind boundary conditions. Initial
data is
\begin{equation}\label{eq:03}
u(x,y,z,0)=U_{in}(x,y,z,0).
\end{equation}
Here, $u_n$ is the outward normal derivative (normal with respect to the boundary  $\partial\Omega_\epsilon$) and $\alpha=\frac{\kappa}{c\rho}$ is the diffusivity. The
heat exchange coefficient $h$ is related to the geometry of the specimen and to the environmental condition close to $S_{top}$. We assume: $h>0$ is constant at $S_{top}$;  $h=0$ on the vertical sides and on the bottom side of the plate.  The incoming heat flux $\phi$ is concentrated at $S_{Bot}$.
Finally, we suppose that  $U^0$ is the same on $S_{Bot}$ and on $S_{Top}$ and that the initial temperature is $U_{in}(x,y,z)=U^0$. 

\subsection{Identification of the background temperature}
\label{ssec:background} 

We have already observed, talking of the flux in subsection  \ref{ssec:ip}, that the only data available in practice are the maps of the bottom side of damaged specimen, recorded by the infrared camera.  In fact we expect that, at least for $t < t_0$, the temperature of $S_{bot}$ is independent of $\epsilon\theta$ with the exception of a small portion of the side just below the damaged area. 
Hence the background temperature in $\Omega_0$ is recovered from the knowledge of $u^\epsilon(x,y,0,t_k)$.\\

If the flux density $\phi=\phi_0$ is constant in space and time (this is the case of {\it stepped heating} or {\it long pulse} thermography), the temperature $u^0=u^0(z,t)$ is independent of $x$ and $y$. Indeed, it can be computed as a series dependent on $z$ and $t$ \cite{CJ59}:

\begin{equation}
u^0(z,t) = \displaystyle \frac{\phi}{h} \left[ 1 + \lambda \left(1 - \frac{z}{a} \right) - \sum_{n=1}^\infty \frac{2 \lambda \left(\beta_n^2 + \lambda^2 \right) cos\left(\beta_n z/a\right)}{\beta_n^2 \left( \lambda + \lambda^2 +\beta_n^2 \right)} e^{-\beta_n^2 \tau} \right]
\label{unidim}
\end{equation}

\noindent
where $\lambda = ah/\kappa$, $\beta_n$ are the real roots of $\beta tan\beta = \lambda$ and $\tau = \alpha t/a^2$.

As the $n_{th}$ root $\beta_n$ belongs to the interval $[(n-1)\pi, (2n-1)\pi/2]$, small times (a few seconds) bring to very large exponents in (\ref{unidim}) for $n > 1$, for the assumed values of the physical and geometrical parameters $\alpha$, $\kappa$, $h$ and $a$. Therefore, the first term of the series (\ref{unidim}) gives a sufficiently good approximation of the background temperature.\\

For low values of $\kappa$ (negligible longitudinal heat transfer) and $t$ (negligible dependence on $h$ before a time $t_a>0$), equation (\ref{unidim}) can be used successfully for recovering $\phi$ from $u^0$. It gives us a complete quantitative information about the heat flux.

\subsection{Main result in 2D: explicit approximation of the damage}

The  Initial Boundary Value Problem (IBVP) defined by (\ref{eq:01}) (\ref{eq:02}) (\ref{eq:03}) is  well posed (see for example \cite[Chapter 2]{Sa08}) and we  refer to it as to the Direct Model  underlying {\bf IP}. On the other hand, in spite of recent uniqueness results by Isakov (see \cite{Is08}), {\bf IP} has a close relation with the Cauchy problem for the heat equation which is severely ill-posed because its solutions are highly unstable (see for example \cite{BBS85}  Chapter 4).\\

First, we linearize the inverse problem by means of Domain Derivative, a technique that accounts for smallness of $\epsilon$ (the size of the damage) with respect to the thickness $a$ of the plate.
Let $u^\epsilon$ denote the unique solution of IBVP (\ref{eq:01}), (\ref{eq:02}), (\ref{eq:03}). We choose this notation to point out the nonlinear dependence of the solution on $\epsilon\theta$. The Domain Derivative $u'$ is defined as  the (Gateaux) derivative of $u^\epsilon$  with respect to the parameter $\epsilon$ in the direction $\theta$ at $\epsilon=0$.\\

It comes from straightforward calculations (reported in detail in \cite{BCFI10}) and linearity arguments that the function $w \equiv \epsilon u'$ ($\forall \epsilon >0$) fulfills the following IBVP in $\Omega_0 \times (0,T]$
 
\begin{equation}\label{eq:1.1}
c\rho w_t=\kappa \Delta w
\end{equation}
with boundary conditions of the form

\begin{align}
\label{eq:1.2}
\kappa w_z(x,a,t)+h w(x,a,t)=\epsilon\theta(x)(\kappa u^0_{zz}(x,a,t)+h u_z^0(x,a,t))-\epsilon\theta_x(x)\kappa u^0_x(x,a,t)  
\end{align}

\begin{equation}\label{eq:1.21}
\kappa w_z(x,0,t)=0
\end{equation}
and initial data
\begin{equation}\label{eq:1.3}
w(x,z,0)=0.
\end{equation}

We rescaled $u'$ by the parameter $\epsilon$ because $\epsilon\theta$ is actually the damage that we must recover and $w(x,0,t)=\epsilon u'(x,0,t)$ is the first order approximation of the thermal contrast $u^\epsilon(x,0,t)-u^0(x,0,t)$ that we actually measure. \\

Then, we use the thinness of the plate itself to expand the unknown and the data in powers of $a$ (see the Appendix for details). In fact, after scaling the  variables $z$, the scaled Domain Derivative $w$ in  $(-\frac{L}{2},\frac{L}{2})\times(0,1)\times(0,T_{max}]$ fulfills the parabolic equation
\begin{equation}
a^2 \frac{w_t}{\alpha} = a^2 w_{xx}+w_{zz} \label{eq:11}
\end{equation}
with boundary conditions
\begin{equation}
\kappa w_z(x,1,t) + a h  w(x,1,t) = a \epsilon\theta(x) (h u^0_z(x,a,t)+ \kappa u^0_{zz}(x,a,t))-a\epsilon\theta_x(x)\kappa u^0_x(x,a,t)  \label{eq:12}
\end{equation}
\begin{equation}
\kappa w_z(x,0,t) = 0\label{eq:13}
\end{equation}
\begin{equation}
w_x(-L/2,z,t)=w_x(L/2,z,t)=0
\end{equation}
(adiabatic conditions on the vertical sides)
and initial data
\begin{equation}
w(x,z,0)=0. \label{eq:14}
\end{equation}

Plugging the formal expansions

\begin{equation}
w=w_0+a w_1+a^2 w_2 +O(a^3)
\end{equation}
\begin{equation}
\theta=\theta_0+a \theta_1+a^2 \theta_2 +O(a^3)
\label{sum_formula}
\end{equation}
\noindent
into the IBVP above and applying perturbation theory, (see Appendix for details)  we obtain an explicit relationship between the unknown coefficients $\theta_j(x,t)$ and the thermal contrast approximated by means of the domain derivative $w(x,0,t)$. If the flux density $\phi$ is constant in space and time, we have
\begin{equation}
\begin{aligned}
\epsilon\theta_0(x,t) ={} & \frac {h w(x,0,t)}{D(t)}\\
\epsilon\theta_1(x,t) ={} & \frac {c\rho {w}_t(x,0,t)-\kappa {w}_{xx}(x,0,t)}{D(t)}\\
\epsilon\theta_2(x,t) ={} & \frac{h}{2\kappa}\frac{c\rho {w_0}_t-\kappa {w_0}_{xx}}{D(t)}\\
\epsilon\theta_3(x,t) ={} & \frac{1}{3}\frac{c\rho {w_2}_t-\kappa {w_2}_{xx}}{D(t)} \\
     = &  \frac{1}{6\alpha}(\epsilon{\theta_1}_t(x,t) D(t) +\epsilon{\theta_1}(x,t) D_t(t))- \frac{1}{6}D\epsilon{\theta_1}_{xx}(x,t).\\
\end{aligned} 
\label{inv_formula}
\end{equation}
where $D(t) = hu^0_z(a,t)+c \rho(a,t) u^0_t(a,t)$.\\

We test our method with synthetic data in section \ref{sec:num} and then we apply it to the reconstruction of a real damage on the hidden face of an aluminum plate (section \ref{ssec:real}). Numerical differentiation of discrete  data is carried out successfully by means of smoothing techniques and local weighted regression \cite{C79}. \\

\section{Test on the inversion formula with synthetic data}\label{sec:num}

Let $\Omega_0$ be the rectangle $[-L/2,L/2]\times [0,a]$. We test the explicit formula (\ref{sum_formula},\ref{inv_formula}) on a damaged boundary $a-\epsilon\theta$ where $\theta(x)$ is one of the following:
\begin{itemize}
\item an unimodal gaussian function: $e^{-\gamma x^2}$ 
\item a rectangular function: $rect(x,d) = H(x + \frac{d}{2}) - H(x - \frac{d}{2})$, defined in terms of the \emph{Heaviside} function $H(x)$
\item a bimodal gaussian function: $A_1 e^{-\gamma_1 (x-x_1)^2} + A_2 e^{-\gamma_2 (x-x_2)^2}$
\end{itemize}
Our data consist of $u^\epsilon(x,0,t)$ for $x \in [-L/2,L/2]$ and $t=0, ...,T_{max}$.

All simulations have been performed by means of COMSOL Multiphysics\textsuperscript{\textregistered}  \cite{COMSOL}, a commercial code solving Partial Differential Equations based on the Finite Element Method (FEM). Apart from the different defect shapes, the model consists of an aluminum plate having a thickness of 40 mm, 
as in the experiment (see Section \ref{ssec:real}). The assumed physical properties of the plate are: thermal conductivity $\kappa = 237$  $W/m^{2}K$, specific heat capacity $c = 900$ $J/kgK$, density $\rho = 2700$  $kg/m^{3}$. A value of the heat exchange coefficient $h$ of $20 W/m^2K$ is chosen to be compatible with the air flow produced by the commercial fan (actually a fume extractor) used in the laboratory experiments, amounting to $115 m^3/min$. The external heat flux $\phi$ is assumed to be $1000 W/m^2$.

\begin{figure}[!ht]
    \subfloat[]{%
      \includegraphics[width=0.5\textwidth]{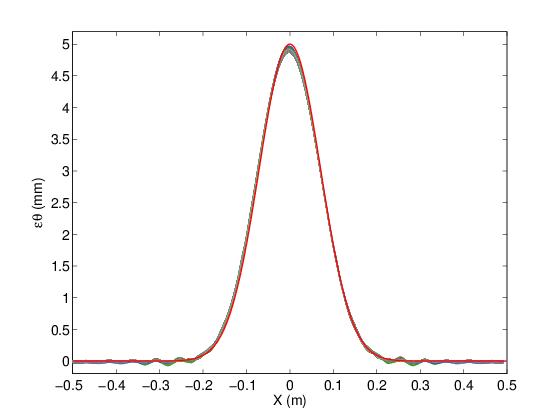}
    }
    \hfill
    \subfloat[]{%
      \includegraphics[width=0.5\textwidth]{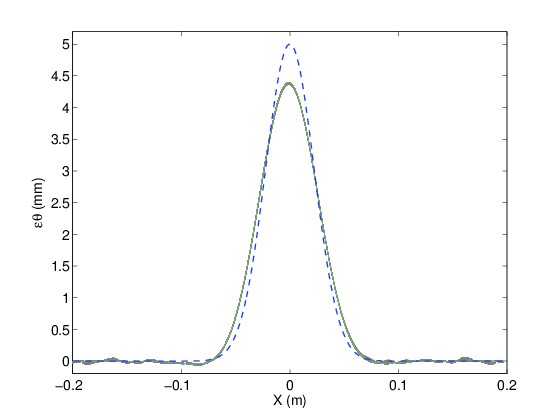}
    }
    \caption{Reconstructed $\epsilon \theta(x)$ for a gaussian defect having $\gamma = 10^2 m^{-2}$ (a) and  $\gamma = 10^3 m^{-2}$ (b), as a function of the observation time}
        \label{fig_gaussian}
\end{figure}

Figure \ref{fig_gaussian} shows the the value of the reconstructed $\epsilon\theta(x)$, computed by means of the inversion formula (\ref{inv_formula}) as a function of time, for different defect widths and $\epsilon = 5 mm$. The reconstruction is time-independent in both cases and we consider this fact a confirm of reliability of the method. While the third-order formula perfectly reconstructs a broad damage ($\gamma = 10^2 m^{-2}$), it appears to slightly underestimate the defect depth for a narrower one ($\gamma = 10^3 m^{-2}$). In both cases the dashed curve represents the actual defect (not visible in figure \ref{fig_gaussian} (a)). It appears that the temperature contrast due to a smaller damage faster deteriorates with time, needing to retain a greater number of terms in the expansion (\ref{inv_formula}).

\begin{figure}[!ht]
    \subfloat[]{%
      \includegraphics[width=0.5\textwidth]{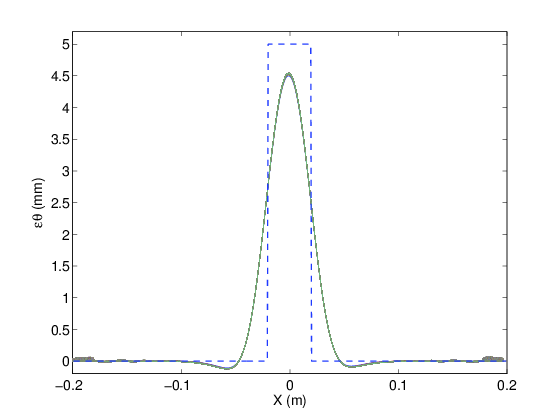}
    }
    \hfill
    \subfloat[]{%
      \includegraphics[width=0.5\textwidth]{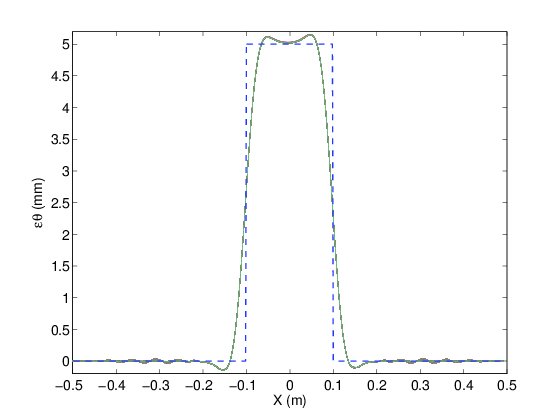}
    }
    \caption{Reconstructed $\epsilon \theta(x)$ for a rectangular defect having $d = 4 cm$ (a) and  $d = 20 cm$ (b), as a function of the observation time}
    \label{fig_rectangular}
\end{figure}

The choice of a gaussian shape is less restrictive than it could seem. Indeed, damage shapes rather different from a gaussian bring to a gaussian-like reconstructions. This is as a consequence of the dissipative character of heat conduction. Something better can be done only if {\it a priori} information about the shape of the damage are available.
Figure \ref{fig_rectangular} shows the computed $\epsilon\theta(x)$ for $\theta = rect(x,d)$ with $d = 4 cm$ (a) and $d = 20 cm$ (b).

\begin{figure}[!ht]
\begin{center}
\includegraphics[width=0.9\textwidth]{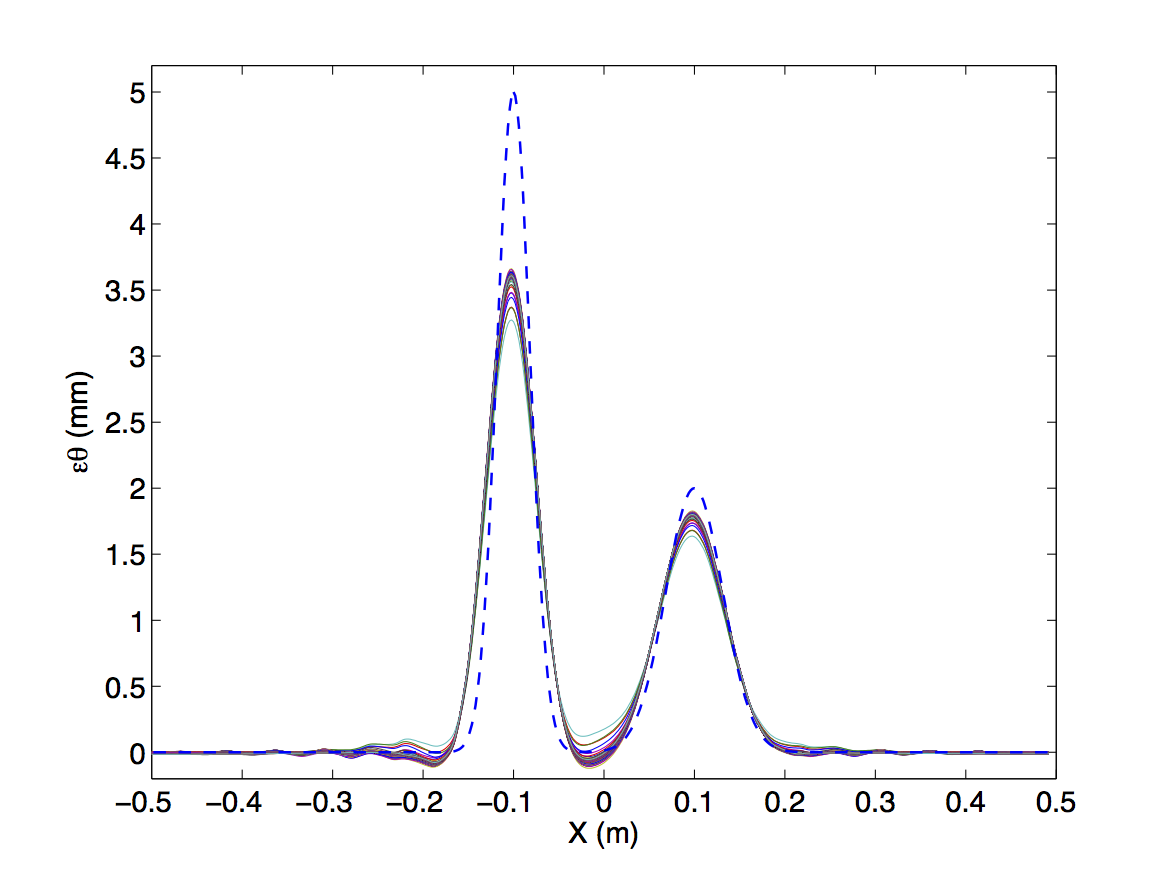}
\end{center}
\caption{Reconstructed $\epsilon \theta(x)$ for a bimodal gaussian damage. The dashed curve represents the actual defect.}
\label{fig_bimodal}
\end{figure}

The inversion method also works well on damages of more complex shape. For example, figure \ref{fig_bimodal} shows the reconstruction of  a defect represented by the overlapping of two gaussian functions: $\epsilon\theta(x) = 5\cdot10^{-3}exp[-1500(x+0.1)^2] + 2\cdot10^{-3}exp[-500(x-0.1)^2]$. As before, the broader damage is quite perfectly reconstructed, although smaller in depth, while the narrower one is underestimated. However, the defect widths are correctly obtained and the inversion formula confirms to be independent on time.

\section{Determination of the heat flux density $\phi$ from thermal data}\label{ssec:phi}

In section \ref{ssec:ip} we already pointed out that the flux density $\phi$, though controlled in some sense, is actually unknown.  Without a reliable estimate of $\phi$ we cannot apply our inversion method to real thermal data.\\

It comes from formula (\ref{unidim}) that a constant flux $\phi_0$ is determined by one temperature measurement taken in the first seconds. Assume now that $\phi$  is a generic smooth positive function so that it can be approximated by the trigonometric polynomial
\begin{equation}
\phi(x) = {\phi_0}+ \sum_{k = 1}^N \left ( \phi_{2k-1} cos\frac{2 k \pi x}{L} + \phi_{2k} sin\frac{2 k \pi x}{L}  \right ).
\label{fourier_sum_fi}
\end{equation}
The coefficients $\phi_j$ for $j=0,1,...,2N$ can be recovered  from the knowledge of the temperature $u(x,0,t)$ measured on the accessible surface $z = 0$. In fact, let $u_{j}(z,t)$ be the solution of the one-dimensional IBVP
\begin{equation}
u_t=\alpha u_{zz}
\end{equation}
\begin{equation}
\kappa u_z(a,t)+h(u(a,t)-U^0)=0
\end{equation}
\begin{equation}
-\kappa u_z(0,t)=\phi_{j}
\end{equation}
\begin{equation}
u(z,0)=U^0.
\end{equation}
There is numerical evidence that for times $t$ such that the ``Fourier number''  $t \alpha/a^2$ is small
\begin{equation}
u(x,0,t) \approx u_0(t) + \sum_{k = 1}^N \left ( u_{2k-1}(t) cos\frac{2 k \pi x}{L} + u_{2k}(t) sin\frac{2 k \pi x}{L}  \right )
\label{unidimensional_sum}
\end{equation}
\noindent
with:
\begin{align}
u_0 & = \frac{\phi_0}{h} \left( 1+ \lambda -\Sigma_N  \right) + U_0 \nonumber \\
u_j & = \frac{\phi_j}{h} \left( 1+ \lambda -\Sigma_N  \right) \label{unidimensional_terms}
\end{align}
\noindent for $j=0,1,...,2N$, where $\lambda$, the $\beta_k$'s and 
\begin{equation}
\Sigma_N(t) = \sum_{k=1}^N \frac{2 \lambda (\beta_k^2 + \lambda^2)}{\beta_k^2(\lambda + \lambda^2 + \beta_k^2)}e^{-\beta_k^2 \alpha t/a^2}
\end{equation}
(a finite truncation of the series appearing in the analytical solution of the unidimensional problem (\ref{unidim})) are defined in \ref{ssec:background}.\\

Since we know $u(x,0,t)$ from our thermal measurements, we can compute the $u_j$'s by means of the scalar products
\begin{align}
u_0(t) & = \frac{1}{L} \int_{-L/2}^{L/2} u(x,t) dx \nonumber \\	
u_{2k-1}(t) & = \frac{1}{L} \int_{-L/2}^{L/2} u(x,t) cos \frac{2 k \pi x}{L} dx \nonumber \\
u_{2k}(t) & = \frac{1}{L} \int_{-L/2}^{L/2} u(x,t) sin \frac{2 k \pi x}{L} dx. \nonumber 
\end{align}

Finally, we invert relations (\ref{unidimensional_terms}) and determine the $\phi_j$'s. In particular, we get
\begin{align}
\phi_0 & =   \frac{(u_0(t) - U_0)h}{1+\lambda+\Sigma_N(t)} \nonumber \\
\phi_{j} & =  \frac{u_j(t) h}{1+\lambda+\Sigma_N(t)}. \\
\end{align}
\noindent

We observe that the "theoretical" $\phi_j$'s are time-independent although they are infinite sums of the time-dependent quantities $u_j(t)$ and $lim_{N \to \infty} \Sigma_N(t)$. We  expect that the numerical approximations produced here are slightly dependent on $ t$ as a consequence of both the truncation of the series and the noise/uncertainty on the temperature data.\\

An independent, alternative way for computing the heat flux would therefore be appreciated as a validation tool.

A simple estimate of $\phi$ can be obtained by measuring the temperature behavior on a thermal insulating plate subjected to the same heat flux density $\phi$. At short times, the low thermal conductivity allows to consider the problem as unidimensional, because the lateral diffusion is negligible and the heating of the plate is nearly adiabatic. The thermal energy $E(t)$ in the specimen, per unit width of the plate, is straightforwardly related to the integral of the temperature at time $t$ over the plate thickness:

\begin{equation}
E(t) = \rho c \int_0^a (u(z,t) - U_0) dz
\end{equation}

\noindent
and the heat flux is computed from the energies at two close times $t_1$ and $t_2$:

\begin{equation}
\phi = \frac{\Delta E}{\Delta t} = \frac{E(t_2) - E(t_1)}{t_2 - t_1}
\end{equation}

Of course, we cannot measure the temperature profile $u(z,t)$ inside the plate. The temperature measured on the accessible surface $z = 0$ , $u(0,t)$ can be used to determine a depth $\delta$ at which $E(t) = \rho c (u(0,t) - U_0) \delta$. Such a depth can easily demonstrated to be related to the so-called \emph{thermal penetration depth}:

\begin{equation}
\delta = \frac{1}{2} \sqrt{\pi \alpha t}
\end{equation}

\begin{figure}[!ht]
\begin{center}
\includegraphics[width=0.9\textwidth]{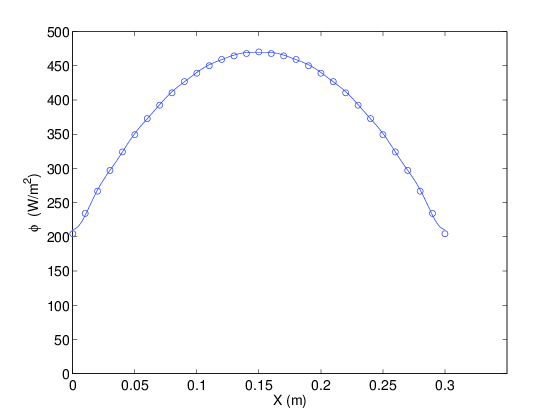}
\end{center}
\caption{Comparison between $\phi$ reconstructed by the Fourier analysis (solid line) and the short-time approximation (circles)}
\label{compare_fourier}
\end{figure}

The above derivation is rigorously true when $\phi$ is constant, independent on $x$. At times $t_1$ and $t_2$ such that $\delta \ll a$, the heat flux is given by:

\begin{equation}
\phi = \frac{\sqrt{\pi}}{2} \left[ \frac{(u(0,t_2) - U_0) \sqrt{t_2} - (u(0,t_1) - U_0) \sqrt{t_1}}{t_2 - t_1} \right] \sqrt{\kappa \rho c}
\label{approx_phi}
\end{equation}

\noindent
the last square root term being the effusivity of the insulating plate. The formula (\ref{approx_phi}) can also be used in the case of a variable flux, $\phi = \phi(x)$, at times short enough to neglect the heat diffusion. Figure \ref{compare_fourier} compares the $\phi$ obtained from the series (\ref{fourier_sum_fi}) with $N = 10$ with the approximation (\ref{approx_phi}) for a true heat flux $\phi(x) = 470 - 1.2\times 10^{4} (x - L/2)^2$. Temperature data have been obtained on a simulated wooden plate of width $L = 1 \, m$ and thickness $a = 8 \, mm$, having thermal parameters $\kappa = 0.12  \, W m^{-1} K^{-1}$, $\rho = 500  \, kg/m^3$, $c = 1760  \, J kg^{-1} K^{-1}$, and assuming $h = 5 \,  W m^{-2} K^{-1}$ on the top surface of the plate.

In section \ref{ssec:real} we will see how the above procedure is applied on real thermographic data.

\section{Recovering the damage from real temperature measurements}\label{ssec:real}

The inversion method described in the previous sections
has been tested on an aluminum plate having a thickness of 10 mm. This aluminum plate is a "thick" metallic object in the sense that it is very hard and far from being a foil. In the meanwhile, from the point of view of mathematical modeling, it is a "thin plate"   because its thickness ($a=10$ mm) is much lower than the size of external surfaces ($a \ll L \approx 200$ mm).\\

Two problems arise when trying to heat a bare metallic surface: 

\begin{enumerate}
\item The very low emissivity (usually lower than 0.1 for a polished aluminum surface) prevents the material to be significantly heated.
\item The high reflectivity makes impossible to take thermographic images of the heated surface, because any object (including the heating lamps) is reflected into the thermographic camera.
\end{enumerate}

The above well-known problems \cite{Ma01} are usually solved by covering the surface with a high-emissivity paint. Such an approach, while suitable in the laboratory, appears to be rather unpractical and usually too much invasive for a real wall. Looking at published emissivity tables  of common materials (e.g. \cite{Ma01}, Table 8.1) matt paper appears to have an emissivity greater than 0.9. That suggested us the possibility of using an adhesive paper sheet to overcome the emissivity/reflectivity problem. 
That approach also modifies the thermal properties of the aluminum plate: the well known analogy among electrical and thermal circuits allows to treat the layered material paper/aluminum as an homogeneous medium having an effective thermal conductivity $\kappa_{e}$ computed as the series of the individual conductivities:

\begin{equation}
\kappa_{e} = \frac{a + d_p} { \frac{d_p}{\kappa_p} + \frac{a}{\kappa}}
\end{equation}

\noindent
where $d_p$ is the paper thickness, $a$ is the thickness of the aluminum plate, $\kappa_p$ and $\kappa$ are the thermal conductivities of paper and aluminum, respectively. For a paper thickness of $8.2 \times 10^{-5} m$, assuming conductivities $\kappa_p = 0.08 \,W m^{-1}K^{-1}$ and $\kappa = 237 \,W m^{-1}K^{-1}$, the effective conductivity is about $9.5 \,W m^{-1}K^{-1}$. The heat capacity and the density are computed according to the same electrical analog model, resulting very close to those of aluminum.

\begin{figure}
\begin{center}
\includegraphics[width=0.9\textwidth]{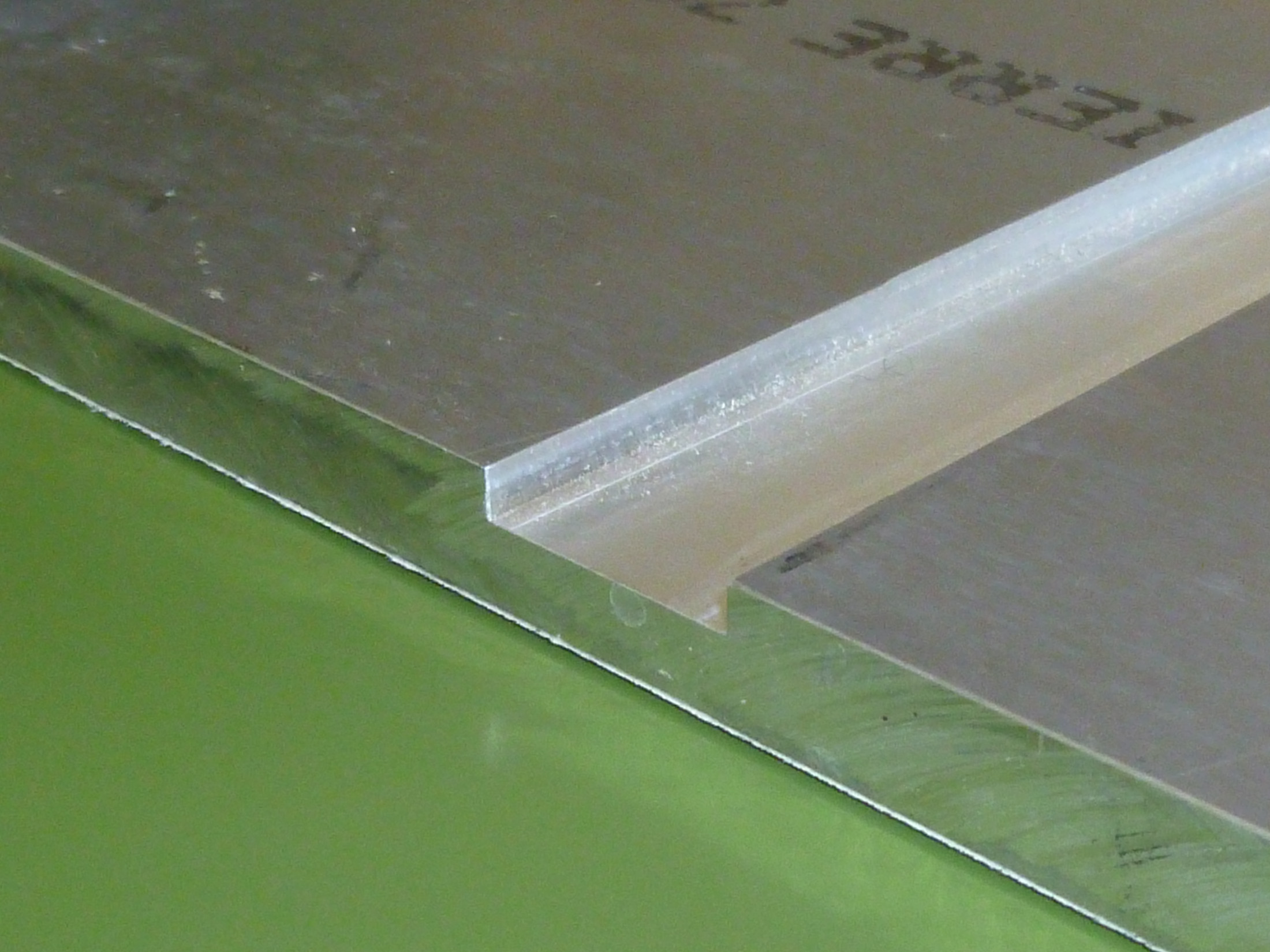}
\end{center}
\caption{The damaged aluminum plate used for the experiment}
\label{plate_foto}
\end{figure}

The damage consists in a rectangular slit of width 20 mm and depth 4 mm, as shown in figure \ref{plate_foto}.

\begin{figure}
\begin{center}
\includegraphics[width=0.9\textwidth]{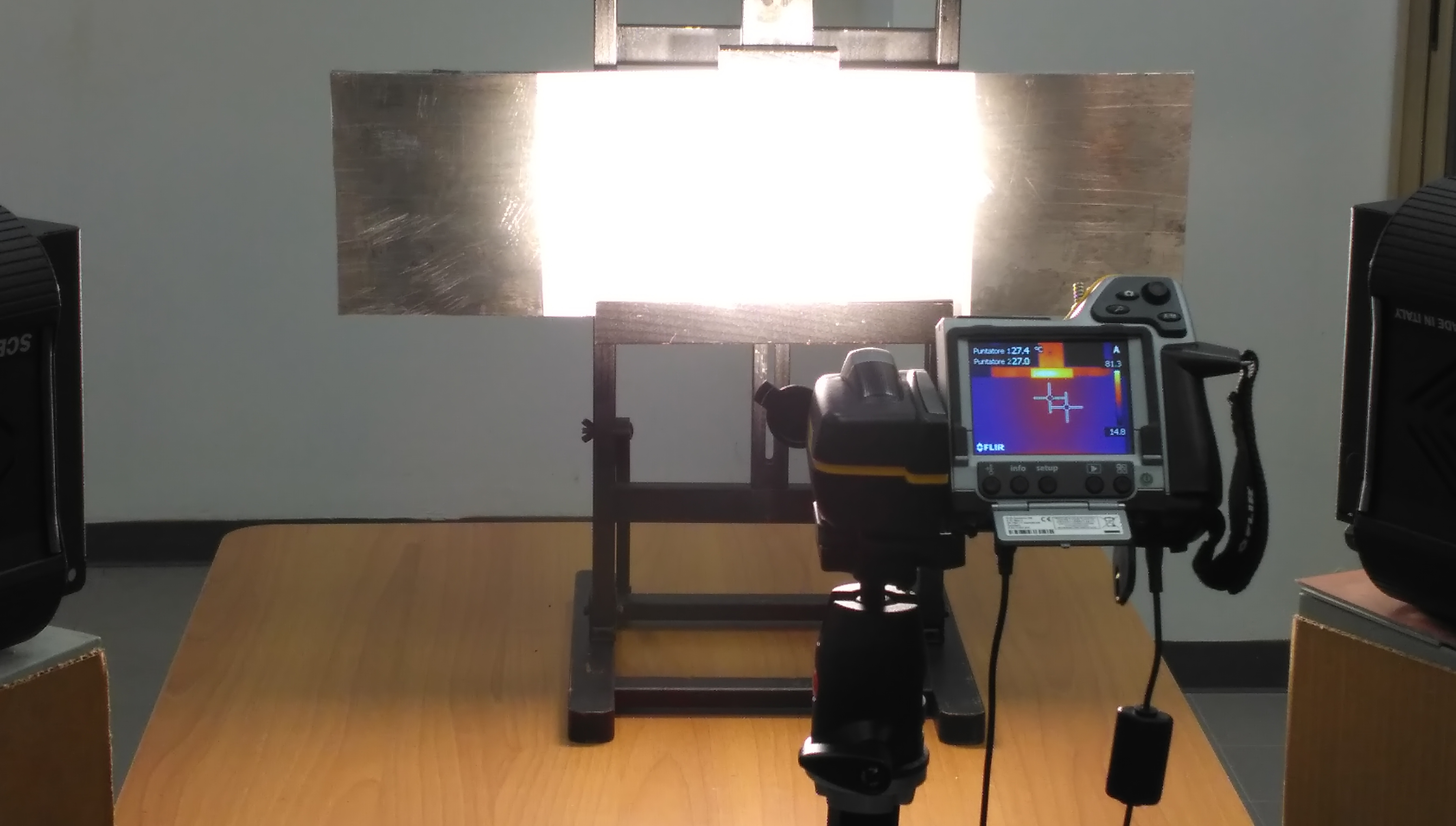}
\end{center}
\caption{The measurement setup}
\label{setup}
\end{figure}

Figure \ref{setup} shows the measurement setup. Two high-power spotlights ($1 kW$ each) are placed and focalized such to produce a high heat flux density on the central region of the plate. The flux, as we will see in the following, has a parabolic shape. A thermographic image is acquired before the heating and for about 400 seconds after power on, at intervals of 5-6 seconds by means of a FLIR B335 thermal camera. 

The thermographic images are successively processed by means of ThermaCAM Researcher\textsuperscript{\textregistered}  software, allowing to acquire the values on a given geometrical region, in particular over a horizontal line parallel to the wide side of the paper sheet, normal to the defect slit.

\subsection{Heat flux estimate}
The heat flux $\phi$ is estimated by taking a short series of thermographic shots on a plywood panel covered with a white matt painting, according to the procedure described in section \ref{ssec:phi}. 

\begin{figure}
\begin{center}
\includegraphics[width=0.9\textwidth]{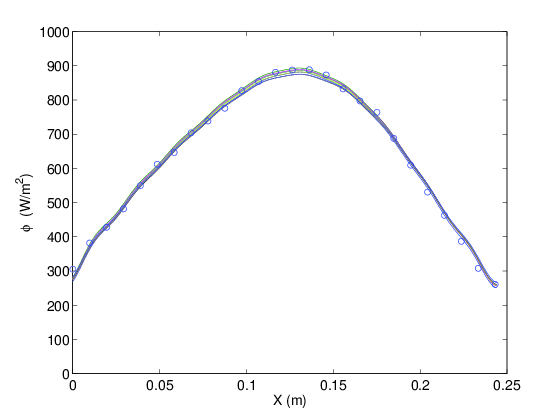}
\end{center}
\caption{Heat flux measured on a plywood panel}
\label{experimental_phi}
\end{figure}

Figure \ref{experimental_phi} compares the heat flux obtained by means of the Fourier analysis (solid lines) at times between 25 and 80 seconds with that computed using the short-time approximation based on the thermal penetration depth. The agreement between the two methods is very satisfactory, and the heat flux  appears to be nearly parabolic in shape. Moreover, the flux is time-independent, as it should.

\begin{figure}
\begin{center}
\includegraphics[width=0.9\textwidth]{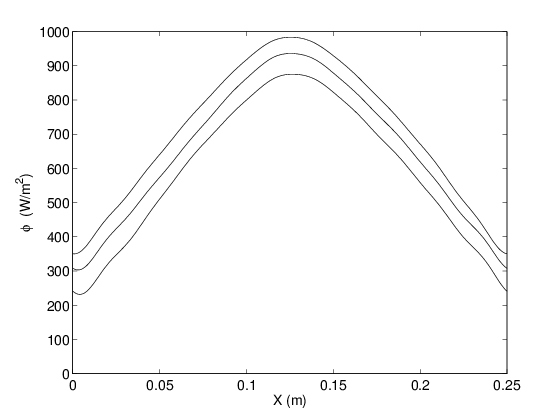}
\end{center}
\caption{Heat flux measured on the aluminum plate with the defect}
\label{exp_phi_lastra}
\end{figure}

The heat flux could also be estimated by the temperature behavior measured on the target aluminum plate, at times small enough to ensure that the influence of the defect on the temperature distribution is negligible. Figure \ref{exp_phi_lastra} shows that the reconstruction of $\phi$ is slightly time-dependent, but it is consistent with that obtained on the insulating panel. The computation based on measurements on a metallic plate is rather sensitive to the \emph{true} value of the thermal conductivity of the material and on the heat exchange coefficient on the top surface.

\subsection{Defect detection}

\begin{figure}
\begin{center}
\includegraphics[width=0.9\textwidth]{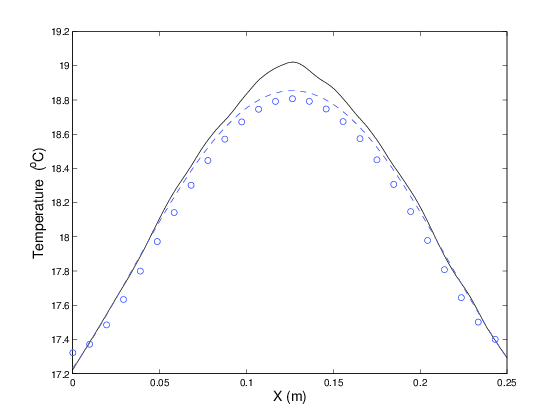}
\end{center}
\caption{Temperature after 1 minute heating (solid line), compared to the fitted background (dashed line) }
\label{temperature_60sec}
\end{figure}

Figure \ref{temperature_60sec} shows the temperature measured on a line crossing the defect, after 60 seconds heating (solid line). The background temperature (dashed line) is computed by fitting a polynomial to the lateral sides of the temperature distribution, i.e. excluding the central portion that is supposedly influenced by the defect on the opposite side. The background temperature could also be computed by simulating the experimental plate with the measured $\phi$ forcing term. That procedure gives reasonable results (circle symbols in figure \ref{temperature_60sec}) but as the computed temperature depends on the assumed physical parameters (thermal characteristics of the plate, heat exchange coefficient at the top surface), the "fitted" background appears to be more reliable and, incidentally, simpler to compute.

\begin{figure}
\begin{center}
\includegraphics[width=0.9\textwidth]{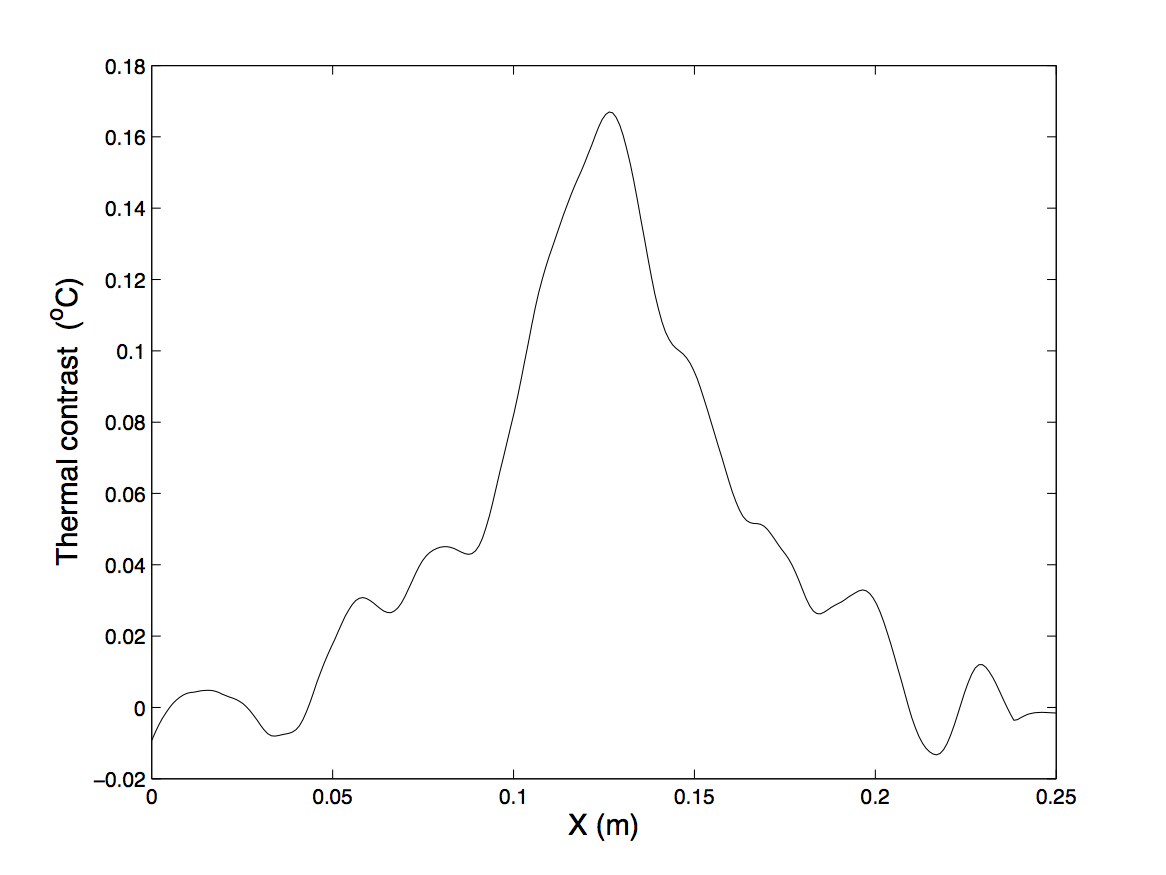}
\end{center}
\caption{Contrast after 1 minute heating}
\label{contrast_60sec}
\end{figure}

The thermal contrast, $u_\epsilon - u_0$ is then computed by subtracting the fitted background  by the measured temperatures at every time. Figure \ref{contrast_60sec} shows the contrast computed at time 60 seconds, from the data of figure \ref{temperature_60sec}. The position of the defect is clearly visible as a peak in the measured contrast.

\begin{figure}
\begin{center}
\includegraphics[width=0.9\textwidth]{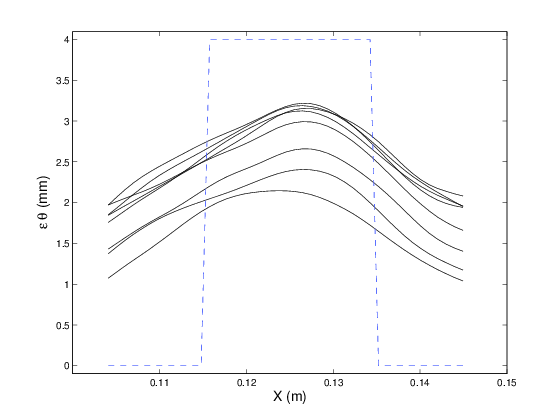}
\end{center}
\caption{Defect (solid lines) reconstructed at times between 30 and 75 seconds, compared to the true defect (dashed line) }
\label{defect}
\end{figure}

The defect shape is recovered by applying the inversion formulas (\ref{sum_formula}, \ref{inv_formula}). Figure \ref{defect} shows that the reconstructed $\epsilon \theta$ (solid lines) slightly depends on time, but nevertheless the correct position and order of magnitude for the true defect (dashed line) is obtained. The shape of the reconstructed defect is gaussian-like as observed with simulated data.

\begin{figure}
\begin{center}
\includegraphics[width=0.9\textwidth]{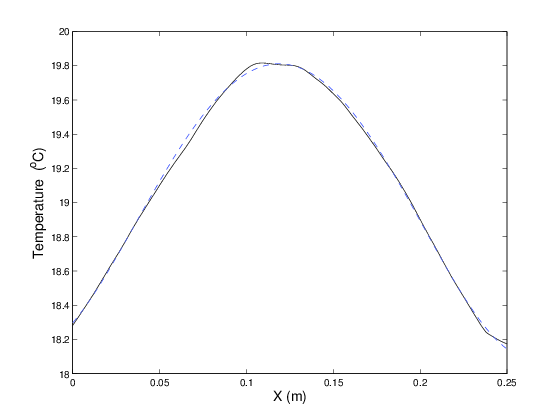}
\end{center}
\caption{Temperature after 1 minute off-center heating (solid line), compared to the fitted background (dashed line) }
\label{temp_dec_60sec}
\end{figure}

\begin{figure}
\begin{center}
\includegraphics[width=0.9\textwidth]{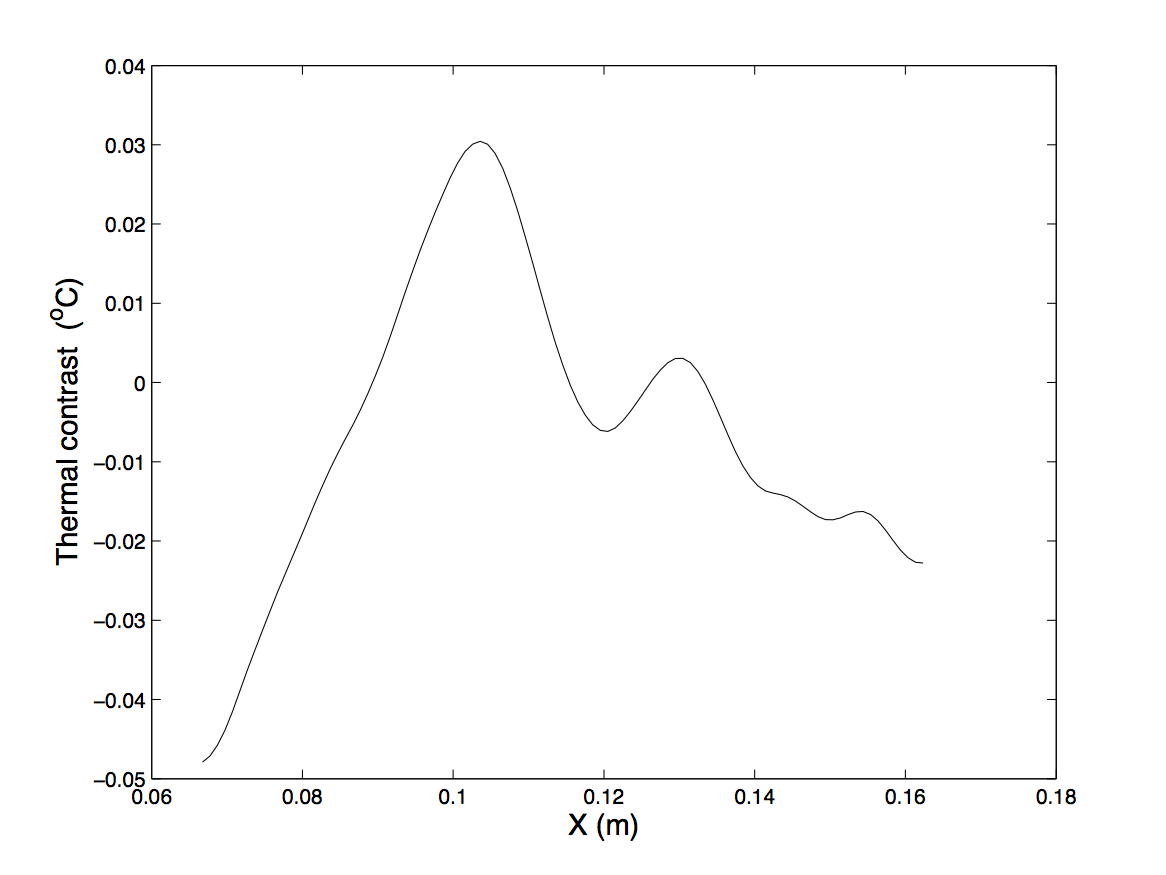}
\end{center}
\caption{Contrast after 1 minute off-center heating}
\label{contrast_dec_60sec}
\end{figure}

The experimental situation shown in figures \ref{temperature_60sec} to \ref{defect} is somehow \emph{ideal}, because the heating maximum is centered on the known position of the defect. A procedure for detecting a defect on a metal plate would be based on scanning the surface by varying the position of the maximum heating. Figure \ref{temp_dec_60sec} shows the temperature measured after 60 seconds heating, when the heating lamps are oriented such to have a maximum at the right of the defect axis. The temperature distribution has a slightly asymmetric shape (perfectly reproduced in numeric simulations), giving rise to a thermal contrast like that shown in figure \ref{contrast_dec_60sec}, where  the  horizontal axis has been limited to a smaller range around the maximum  to emphasize the appearance of an asymmetrical smaller peak at the right of the defect, due to the heating maximum.

\begin{figure}
\begin{center}
\includegraphics[width=0.9\textwidth]{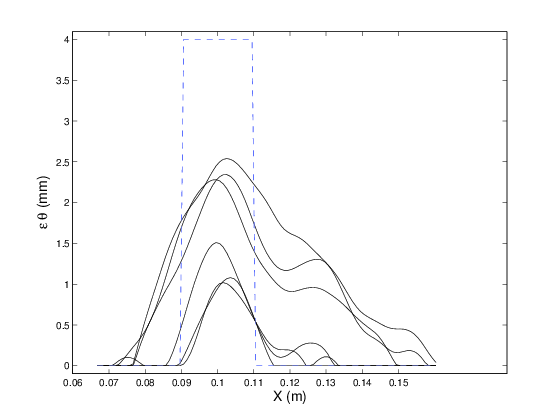}
\end{center}
\caption{Defect (solid lines) reconstructed at times between 30 and 75 seconds for off-center heating, compared to the true defect (dashed line) }
\label{defect_dec}
\end{figure}

The reconstruction of the defect is however possible also with an off-center illumination, as figure \ref{defect_dec} shows. The appearance of a spurious \emph{tail} on the right of the true defect is due to the \emph{wrong} illumination, but the size of the defect is properly obtained. 

\section*{Appendix. Thin Plate Approximation and Perturbation Theory. The special case of two-dimensional domain and flux $\Phi$ constant in $x$ e $t$}

\begin{figure}[!ht]
\begin{center}
\includegraphics[width=0.5\textwidth,trim={0 200 0 0},clip]{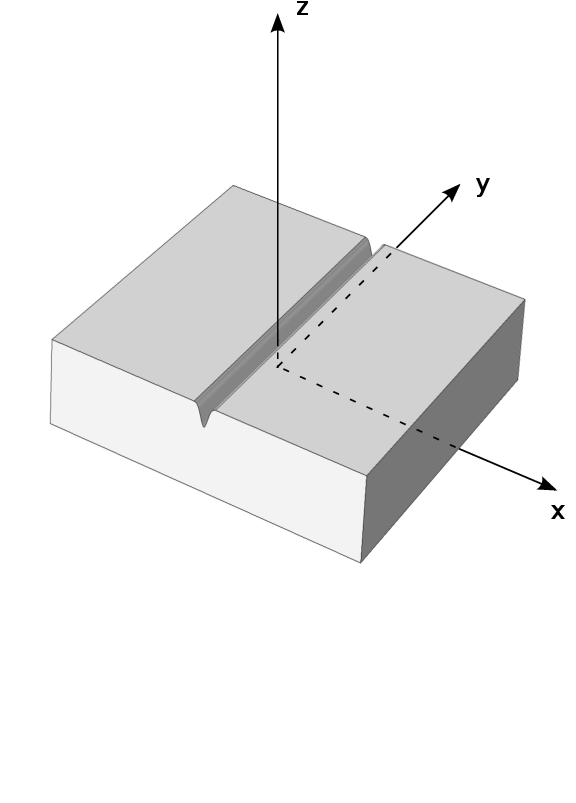}
\end{center}
\caption{A "cylindrical" geometry justifying the 2D approach}
\label{model3D}
\end{figure}

If the  function $\theta$ is "cylindrical" (i.e. $\theta(x,y)=\theta(x)$ $\forall y \in [-\frac{L}{2},\frac{L}{2}]$),  for large values of $L$,  problem {\bf IP'} is essentially a two dimensional one. It happens, in particular, when the defected region unevenly develops along one of the coordinates, as shown in figure \ref{model3D}.
In this section, we use  2D  Thin Plate Approximation (TPA) (see \cite{KSV94}, \cite{In97}, \cite{IO16}) to build an approximate explicit formula for the reconstruction of the damage $\epsilon\theta$.\\

If we have no \emph{a priori} information about the localization of the damage it is natural to use  a heat flux constant in both space and time variables.  Hence, the background temperature $u^0$ is independent of $x$ and will be written as $u^0(z,t)$.  In this case, inversion formulas turns out to be much simpler that the general 2D and 3D cases reported in Appendix. Although, it is not easy to produce a spatially constant heat flux, we will see in the numerical section \ref{sec:num} that  "equivalent" constant fluxes give very accurate reconstructions.\\

First of all, we change the  variable $z$ so that the thickness of the domain remains equal to one when $a$ goes to zero in the Thin Plate Approximation:  we rescale $z$ and transform it in the new variable  $\zeta=\frac{z}{a}$. \\

Since $w_\zeta=\frac{w_z}{a}$ we get the following IBVP for domain derivative

$$\frac{a^2}{\alpha} w_t = a^2 w_{xx}+w_{\zeta\zeta}$$
$$\kappa w_\zeta(x,1,t) + a h  w(x,1,t) = a \epsilon\theta(x) (h u^0_z(a,t)+c\rho u^0_t(a,t))
$$
$$\kappa w_\zeta(x,0,t) = 0 $$
$$w_x(-L/2,\zeta,t)=w_x(L/2,\zeta,t)=0$$ (i.e. adiabatic conditions on the vertical sides) and
$w(x,\zeta,0)=0$ (initial condition).\\

Moreover, we have the extra boundary condition
$$w(x,0,t) \approx u^\epsilon(x,0,t)-u^0(0,t)$$
obtained in practice by means of an infrared camera and necessary for the reconstruction of $\epsilon\theta$. The notation
$$D(t)=h u^0_z(a,t)+c\rho u^0_t(a,t)$$
will be useful in what follows.

We expand formally $w$ and $\theta$
$$w=w_0+a w_1+a^2 w_2 +O(a^3)$$
$$\theta=\theta_0+a \theta_1+a^2 \theta_2 +O(a^3),$$
and plug the expansions into the IBVP. \\

\noindent {\it Order zero}
We have
$${w_0}_{\zeta\zeta}=0$$
with ${w_0}_\zeta(x,1,t)={w_0}_\zeta(x,0,t)=0$
so that $w_0(x,\zeta,t)=w_0(x,0,t)$ for all $\zeta \in [0,1]$.\\

\noindent {\it Order one}
We have
$${w_1}_{\zeta\zeta}=0$$
with 
$$\kappa {w_1}_\zeta(x,1,t)+hw^0(x,1,t)=\epsilon\theta_0(x)D(t)$$
and 
 $${w_1}_\zeta(x,0,t)=0.$$
Since ${w_1}_\zeta(1)={w_1}_\zeta(0)+\int_0^1{w_1}_{\zeta\zeta}(\zeta)d\zeta=0$, we have that also $w_1$ is independent of $\zeta$ and
\begin{equation}
\epsilon\theta_0(x,t)= \frac{h w^0(x,t)}{D(t)}.\label{teta0}
\end{equation}
\\

\noindent {\it Order two}\\

We have
$${w_2}_{\zeta\zeta}=\frac{{w_0}_t}{\alpha}-{w_0}_{xx}.$$
Since $w^0$ is independent on $\zeta$, we have
\begin{equation}
w_2(x,\zeta,t)=(\frac{{w_0}_t(x,t)}{\alpha}-{w_0}_{xx}(x,t)) \frac{\zeta^2}{2}.\label{w2}
\end{equation}

Observe that, from (\ref{w2}) we have $w_2(x,0,t) = 0$ for all $x$ and $t$. For $n \ge 2$ we have
\begin{equation}
{w_{n+1}}_{\zeta\zeta}=\frac{{w_{n-1}}_t}{\alpha}-{w_{n-1}}_{xx}=G_{n-1}(x,t) \frac{\zeta^{n-1}}{(n-1)!}\label{w11}
\end{equation}
so that, for $n \ge 2$
\begin{equation}
w_n(x,0,t) \equiv 0.\label{maggn}
\end{equation}

Hence, 
$$w_0(x,0,t)+aw_1(x,0,t)=w(x,0,t) \approx u^\epsilon(x,0,t)-u^0(x,0,t).$$
A setting of $w^0$ and $w^1$ compatible with (\ref{maggn}) is
$w_0(x,0,t)=w(x,0,t)$ and $w_1(x,0,t) \equiv 0$. It means that $w_1(x,\zeta,t) \equiv 0$ in $\Omega_0 \times (0,T_{max}]$. Also $w_3\equiv 0$ (from (\ref{w11}). It is easy to prove by induction that $w_{2n+1}\equiv 0$ $\forall n=1,2,3,...$.\\

Hence, on the top boundary of the rectangle, we have 
$$\kappa {w_2}_\zeta=\epsilon \theta_1 D(t)=\kappa \int_0^1 (\frac{{w_0}_t}{\alpha}-{w_0}_{xx})dx$$
and, consequently,
\begin{equation}
\epsilon\theta_1(x,t)= \frac{c\rho {w_0}_t(x,t)-\kappa {w_0}_{xx}(x,t)}{D(t)}.\label{teta1}
\end{equation}
\\

\noindent {\it Order three}\\

We have
\begin{equation}
\kappa {w_3}_\zeta(x,1,t)+h w_2(x,1,t)=\epsilon \theta_2(x) D(t)
\end{equation}
Since $w_3$ vanishes everywhere we have
\begin{equation}
\epsilon\theta_2(x,t)= \frac{h}{2\kappa}\frac{c\rho {w_0}_t-\kappa {w_0}_{xx}}{D}.
\end{equation}
\\

\noindent {\it Order four}\\

Since
\begin{equation}
\kappa {w_4}_\zeta(x,1,t)+h w_3(x,1,t)=\epsilon \theta_3(x) D(t)
\end{equation}
we have
\begin{equation}
\epsilon \theta_3(x,t) D(t) = \kappa \int_0^1 (\frac{{w_2}_t(x,t)}{\alpha}-{w_2}_{xx}(x,t))dz.
\end{equation}
Since
\begin{equation}
\frac{{w_0}_t(x,t)}{\alpha}-{w_0}_{xx}(x,t))=\frac{D(t)\epsilon\theta_1(x)}{\kappa}
\end{equation}
we have $w_2(x,\zeta,t)=\frac{D(t)\epsilon\theta_1(x,t)}{\kappa}\frac{\zeta^2}{2}$ 
so that
\begin{equation}
\epsilon\theta_3(x,t) = \frac{1}{6\alpha}(\epsilon{\theta_1}_t(x,t) D(t) +\epsilon{\theta_1}(x,t) D_t(t))- \frac{1}{6}D\epsilon{\theta_1}_{xx}(x,t).
\end{equation}
\\
\noindent{\it Remark about non constant flux}
Although in practice we can use a constant "equivalent" flux density (see section \ref{ssec:real}), 
equations for $\theta_k(x,t)$ can be written also for $\phi=\phi(x)$. In this case, instead of algebraic 
relations like (\ref{teta0}) or (\ref{teta1}) we have ordinary differential equations of the first order, due to 
the presence of the derivative $\theta_x$ in the boundary conditions of domain derivative (see 
(\ref{eq:1.2}).

\begin{figure}[!ht]
\begin{center}
\includegraphics[width=0.8\textwidth]{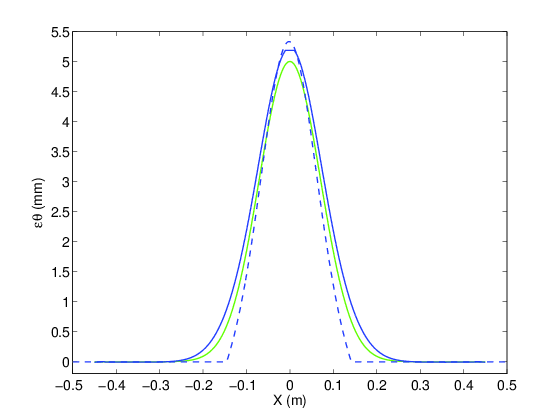}
\end{center}
\caption{Damage reconstruction in the variable flux case}
\label{eq_diff}
\end{figure}

If we are modeling the full dimensional problem in which $\phi=\phi(x,y)$, the damage 
$\epsilon\theta$ solves a linear partial differential equation of the first order. These differential 
equations can be solved analytically, but they require a special care in the neighborhood of possible 
maxima of the temperature where the equation cannot be written in normal form. Although this is still a 
work in progress, in figure \ref{eq_diff} we show the reconstruction of a synthetic damage (green line) 
detected in 2D with a non constant flux solving a ODE in singular implicit form (continuous blue line) 
compared with the solution obtained with constant equivalent flux density (dashed line).

\section{Conclusions}

In this paper we derive formally an explicit approximation of the solution of a nonlinear inverse problem for the heat equation on a thin conducting plate. More precisely, we recover an inaccessible surface damage $\epsilon\theta$ from thermal data.  First we linearize the direct model by means of Domain Derivative. Then, we use perturbation theory in order to construct the Thin Plate Approximation of $\epsilon\theta$ at the first order. Finally, we test the reconstruction strategy with real thermal data.\\

\end{document}